\documentclass[11pt]{article}
\usepackage{amsfonts}
\usepackage{dsfont}
\usepackage{amssymb}
\usepackage{amsmath}
\usepackage{amsxtra}
\usepackage{graphicx}
\usepackage{psfrag}
\usepackage{mathrsfs}
\usepackage{multirow}
\usepackage[sort, numbers, compress]{natbib}
\usepackage{stmaryrd}
\usepackage{subfigure}
\usepackage{tikz}
\usepackage{empheq}
\usepackage[normalem]{ulem}
\usepackage{pdfpages}
\usepackage{color}
\usepackage{longtable}
\usepackage{footmisc}
\usepackage{hyperref}

\usepackage{lineno}

\begin{document}
\def\BGamma{\mbox{\boldmath$\Gamma$}}
\def\BDelta{\mbox{\boldmath$\Delta$}}
\def\BTheta{\mbox{\boldmath$\Theta$}}
\def\BLambda{\mbox{\boldmath$\Lambda$}}
\def\BXi{\mbox{\boldmath$\Xi$}}
\def\BPi{\mbox{\boldmath$\Pi$}}
\def\BSigma{\mbox{\boldmath$\Sigma$}}
\def\BUpsilon{\mbox{\boldmath$\Upsilon$}}
\def\BPhi{\mbox{\boldmath$\Phi$}}
\def\BPsi{\mbox{\boldmath$\Psi$}}
\def\BOmega{\mbox{\boldmath$\theta$}}
\def\Balpha{\mbox{\boldmath$\alpha$}}
\def\Bbeta{\mbox{\boldmath$\beta$}}
\def\Bgamma{\mbox{\boldmath$\gamma$}}
\def\Bdelta{\mbox{\boldmath$\delta$}}          
\def\Bepsilon{\mbox{\boldmath$\epsilon$}}
\def\Bzeta{\mbox{\boldmath$\zeta$}}
\def\Beta{\mbox{\boldmath$\eta$}}
\def\Btheta{\mbox{\boldmath$\theta$}}
\def\Biota{\mbox{\boldmath$\iota$}}
\def\Bkappa{\mbox{\boldmath$\kappa$}}
\def\Blambda{\mbox{\boldmath$\lambda$}}
\def\Bmu{\mbox{\boldmath$\mu$}}
\def\Bnu{\mbox{\boldmath$\nu$}}
\def\Bxi{\mbox{\boldmath$\xi$}}
\def\Bpi{\mbox{\boldmath$\pi$}}
\def\Brho{\mbox{\boldmath$\rho$}}
\def\Bsigma{\mbox{\boldmath$\sigma$}}
\def\Btau{\mbox{\boldmath$\tau$}}
\def\Bupsilon{\mbox{\boldmath$\upsilon$}}
\def\Bphi{\mbox{\boldmath$\phi$}}
\def\Bchi{\mbox{\boldmath$\chi$}}
\def\Bpsi{\mbox{\boldmath$\psi$}}
\def\Bomega{\mbox{\boldmath$\omega$}}
\def\Bvarepsilon{\mbox{\boldmath$\varepsilon$}}
\def\Bvartheta{\mbox{\boldmath$\vartheta$}}
\def\Bvarpi{\mbox{\boldmath$\varpi$}}
\def\Bvarrho{\mbox{\boldmath$\varrho$}}
\def\Bvarsigma{\mbox{\boldmath$\varsigma$}}
\def\Bvarphi{\mbox{\boldmath$\varphi$}}
\def\bone{\mbox{\boldmath$1$}}
\def\bzero{\mbox{\boldmath$0$}}
\def\bnabla{\mbox{\boldmath$\nabla$}}
\def\bvarepsilon{\mbox{\boldmath$\varepsilon$}}
\def\bA{\mbox{\boldmath$ A$}}
\def\bB{\mbox{\boldmath$ B$}}
\def\bC{\mbox{\boldmath$ C$}}
\def\bD{\mbox{\boldmath$ D$}}
\def\bE{\mbox{\boldmath$ E$}}
\def\bF{\mbox{\boldmath$ F$}}
\def\bG{\mbox{\boldmath$ G$}}
\def\bH{\mbox{\boldmath$ H$}}
\def\bI{\mbox{\boldmath$ I$}}
\def\bJ{\mbox{\boldmath$ J$}}
\def\bK{\mbox{\boldmath$ K$}}
\def\bL{\mbox{\boldmath$ L$}}
\def\bM{\mbox{\boldmath$ M$}}
\def\bN{\mbox{\boldmath$ N$}}
\def\bO{\mbox{\boldmath$ O$}}
\def\bP{\mbox{\boldmath$ P$}}
\def\bQ{\mbox{\boldmath$ Q$}}
\def\bR{\mbox{\boldmath$ R$}}
\def\bS{\mbox{\boldmath$ S$}}
\def\bT{\mbox{\boldmath$ T$}}
\def\bU{\mbox{\boldmath$ U$}}
\def\bV{\mbox{\boldmath$ V$}}
\def\bW{\mbox{\boldmath$ W$}}
\def\bX{\mbox{\boldmath$ X$}}
\def\bY{\mbox{\boldmath$ Y$}}
\def\bZ{\mbox{\boldmath$ Z$}}
\def\ba{\mbox{\boldmath$ a$}}
\def\bb{\mbox{\boldmath$ b$}}
\def\bc{\mbox{\boldmath$ c$}}
\def\bd{\mbox{\boldmath$ d$}}
\def\be{\mbox{\boldmath$ e$}}
\def\bff{\mbox{\boldmath$ f$}}
\def\bg{\mbox{\boldmath$ g$}}
\def\bh{\mbox{\boldmath$ h$}}
\def\bi{\mbox{\boldmath$ i$}}
\def\bj{\mbox{\boldmath$ j$}}
\def\bk{\mbox{\boldmath$ k$}}
\def\bl{\mbox{\boldmath$ l$}}
\def\bm{\mbox{\boldmath$ m$}}
\def\bn{\mbox{\boldmath$ n$}}
\def\bo{\mbox{\boldmath$ o$}}
\def\bp{\mbox{\boldmath$ p$}}
\def\bq{\mbox{\boldmath$ q$}}
\def\br{\mbox{\boldmath$ r$}}
\def\bs{\mbox{\boldmath$ s$}}
\def\bt{\mbox{\boldmath$ t$}}
\def\bu{\mbox{\boldmath$ u$}}
\def\bv{\mbox{\boldmath$ v$}}
\def\bw{\mbox{\boldmath$ w$}}
\def\bx{\mbox{\boldmath$ x$}}
\def\by{\mbox{\boldmath$ y$}}
\def\bz{\mbox{\boldmath$ z$}}
\newcommand{\kg}[1]{\textcolor{blue}{\textbf{(kg)} #1}}
\newcommand{\gt}[1]{\textcolor{red}{\textbf{(gt)} #1}}
\newcommand{\zw}[1]{\textcolor{purple}{\textbf{(zw)} #1}}
\newcommand{\xz}[1]{\textcolor{brown}{\textbf{(xz)} #1}}
\newcommand{\mct}[1]{\textcolor{green}{\textbf{(mct)} #1}}
\title{
System inference via field inversion for the spatio-temporal progression of infectious diseases: Studies of COVID-19 in Michigan and Mexico}
\author{Z.~Wang, M.~Carrasco-Teja, X.~Zhang, G.H.~Teichert \& K.~Garikipati\thanks{Corresponding author, \texttt{krishna@umich.edu}}\\Mechanical Engineering, Mathematics and the Michigan Institute for Computational \\Discovery \& Engineering, University of Michigan}

\maketitle
\begin{abstract}
We present an approach to studying and predicting the spatio-temporal progression of infectious diseases. We treat the problem by adopting a partial differential equation (PDE) version of the Susceptible, Infected, Recovered, Deceased (SIRD) compartmental model of epidemiology, which is achieved by replacing compartmental populations by their densities. Building on our recent work (\emph{Computational Mechanics}, \textbf{66}, 1177, 2020), we replace our earlier use of global polynomial basis functions with those having local support, as epitomized in the finite element method, for the spatial representation of the SIRD parameters. The time dependence is treated by inferring constant parameters over time intervals that coincide with the time step in semi-discrete numerical implementations. In combination, this  amounts to a scheme of field inversion of the SIRD parameters over each time step. Applied to data over ten months of 2020 for the pandemic in the US state of Michigan and to all of Mexico, our system inference via field inversion infers spatio-temporally varying PDE SIRD parameters that replicate the progression of the pandemic with high accuracy. It also produces accurate predictions, when compared against data, for a three week period into 2021. Of note is the insight that is suggested on the spatio-temporal variation of infection, recovery and death rates, as well as patterns of the population's mobility revealed by diffusivities of the compartments.

\end{abstract}
\section{Introduction}
\label{sec:background}

Classical epidemiological models, such as the Susceptible-Infected-Recovered (SIRD) model~\cite{Kermack1927}, are ordinary differential equations (ODEs) defined by specifying the compartmental sub-population numbers over some geographical region. Spatial effects have typically been introduced by resolving smaller regions and treating them individually\cite{Eisenberg2015,Eisenberg2013-SIWRCholera,Wesoloski2012-Malaria,Colizza2007-InfluenzaMetapopulations,hethcote2000mathematics}. During the long-lasting and widespread epidemics, such as the COVID-19 Pandemic, the effects on the infection rate of imposing--and then lifting--mobility restrictions and social distancing mandates revolve on the question of the time and spatially varying mobility of the population. At the finest resolution, this must be approached via agent-based models\cite{Hunter2017-ABMReview}, using individuals' mobility data. However, this data is not available for the entire population, and contact tracing campaigns face challenges of recruiting workers, access, technology, as well as socio-political resistance. Against these difficulties, an intriguing question to explore is whether simple reaction-diffusion models can detect the evidence of mobility in these data. Such an approach must start with a partial differential equation (PDE) version of the epidemiological models, which is easily defined by converting compartmental sub-populations to densities over sub regions by normalizing with the corresponding areas. To address the mobility of the population, diffusion terms are introduced to the SIRD model, which is transformed to a set of reaction-diffusion PDEs in two spatial dimensions \cite{Viguerie2020,Zohdi2020,Chang2020,Wangetal-COVID2020}. 

The widespread availability of data in the public domain \cite{1point3acres,yang2020covidnet,jhumap,MI-covid19-data,NYT-data,IHME-data,Inegi-mex,conacyt-mex} has spurred a widespread interest among computational and data scientists, who have sought to test and refine their methods against these repositories. This has opened up the possibility that advances in computational and data science may contribute to the existing and rapidly expanding body of work in epidemiology, in inferring the dynamics of COVID-19 and making projections. We have similarly sought to build off our recent work in data-driven and machine learning approaches~\cite{WangCMAME2019,WangWu2020,WangCMAME2021,Wang2020-brain,Wang2020-MRu,Teichert2018,Teichert2019,Teichert2020,Zhang2020,Zhang2021mlpde} and presented a class of system identification techniques for inference of ODE and PDE forms of the SIRD model, as well as Bayesian neural networks for representation and uncertainty quantification-guided prediction \cite{Wangetal-COVID2020}. That work focused on the US state of Michigan. In this communication, we revise our approach for inference of the PDE SIRD model with \emph{temporal} and \emph{spatial} evolving parameters and diffusivities. Importantly, instead of global polynomial representations of PDE SIRD parameters over the spatial and temporal domains, we adopt field inversion over time intervals that coincide with the time steps of our underlying numerical implementation. This affords much greater accuracy over the global polynomial ansatz. Adjoint-based gradient optimization for field inversion of parameters at each time step replaces the use of stepwise regression-based system identification in our previous work. We find that the improved accuracy with respect to the data over the time interval of inference, as well as of the predictions, is worth the increased expense. We have brought abundant high-quality, public domain, data \cite{1point3acres,yang2020covidnet,jhumap,MI-covid19-data,NYT-data,IHME-data,Inegi-mex,conacyt-mex} on the evolution of COVID-19 in both, the State of Michigan, with a population of 9.98 million, distributed in 83 counties, over 250,493 km$^2$, and the country of Mexico, with a population of 126 million, distributed in 32 geographical entities (31 states, plus Mexico City), on 1,972,550 km$^2$. The temporal resolution by days and spatial resolution by counties/states have allowed us to study the mobility in these data using our methods of system inference.

In Section \ref{sec:SIRDmodel} we review our previous work of system inference for the spatio-temporal SIRD model first, and then extend it by incorporating temporal and spatial parameters and diffusivities using a finite element representation. The PDE SIRD model-constrained inference are presented in Section \ref{sec:adjoint}. Section \ref{sec:data} is on data preparation. The results for inference of classical SIRD parameters as well as the diffusivities, and forward prediction are presented in Section \ref{sec:results}. Our conclusions appear in Section \ref{sec:concl}.

\section{Compartmental differential equations models of infectious disease dynamics}\label{sec:SIRDmodel}

We begin with the conventional SIRD compartmental epidemiology model. The population, taken to remain constant at $N$, is divided into four disjoint compartments with time-dependent sub-populations: $S(t)$ for susceptible, $I(t)$ for infected, $R(t)$ for recovered and  $D(t)$ for deceased individuals. The governing system of ordinary differential equations (ODEs) is:
\begin{align}
    \frac{\text{d} S}{\text{d} t} &=-\frac{\beta}{N} SI+\gamma R\label{eq:S}\\
    \frac{\text{d}I}{\text{d}t} &=\frac{\beta}{N}SI-\mu I-\alpha I\label{eq:I}\\
    \frac{\text{d}R}{\text{d}t} &=\mu I-\gamma R\label{eq:R}\\
    \frac{\text{d}D}{\text{d}t} &=\alpha I\label{eq:D}\\
    N &= S(t) + I(t) + R(t) + D(t).\label{eq:N}
\end{align}

This is the canonical form of the model where the sub-populations are assumed to be well-mixed so that spatial variations can be ignored over the domain of interest. Here $\beta$ is the infection rate, $\mu$ is the recovery rate, $\gamma$ is the rate of immunity loss, and $\alpha$ is the death rate. 

We have extended the SIRD model to a system of partial differential equations (PDEs) in two spatial dimensions using the same compartments~\cite{Wangetal-COVID2020}. However, the population variables are now replaced with spatio-temporally varying densities, $\widehat{S}(\boldsymbol{x},t),\widehat{I}(\boldsymbol{x},t),\widehat{R}(\boldsymbol{x},t),\widehat{D}(\boldsymbol{x},t)$ defined as numbers per unit area.

\begin{align}
    \frac{\partial \widehat{S}}{\partial t} &=\mathcal{D}_\text{S}\nabla^2 \widehat{S}-\frac{\beta}{\widehat{N}} \widehat{S}\widehat{I}+\gamma \widehat{R}\label{eq:S_2D}\\
    \frac{\partial \widehat{I}}{\partial t} &=\mathcal{D}_\text{I}\nabla^2 \widehat{I}+\frac{\beta}{\widehat{N}}\widehat{S}\widehat{I}-\mu \widehat{I}-\alpha \widehat{I}\label{eq:I_2D}\\
    \frac{\partial\widehat{R}}{\partial t} &=\mathcal{D}_\text{R}\nabla^2 \widehat{R}+\mu \widehat{I}-\gamma \widehat{R}\label{eq:R_2D}\\
    \frac{\partial\widehat{D}}{\partial t} &=\alpha \widehat{I}\label{eq:D_2D}
\end{align}
where $\mathcal{D}_\text{S}, \mathcal{D}_\text{I}, \mathcal{D}_\text{R}$ are diffusivities of the corresponding  compartments, and represent the mobility of the sub-population via random walks. We define $\widehat{(\bullet)}=(\bullet)/\int_{\Omega}\text{d}A$ where $\Omega$ is the domain of study: either the lower peninsula of the State of Michigan,  or the territory of the country of Mexico. Furthermore the population constraint holds: $\int_{\Omega}\widehat{N}\text{d}A = \int_{\Omega}\widehat{S}(t)\text{d}A  + \int_{\Omega}\widehat{I}(t)\text{d}A + \int_{\Omega}\widehat{R}(t)\text{d}A + \int_{\Omega}\widehat{D}(t)\text{d}A$.
In what follows of this communication, we only consider the PDEs SIRD model, and, for the sake of readability, we dispense with the hats on the compartments.

We adopt the weak form, and specifically, the finite element framework for the above system of PDEs. For a generic, finite-dimensional field $u^h$, the problem is stated as follows: find $u^h\in \mathscr{S}^h \subset \mathscr{S}$, where $\mathscr{S}^h= \{ u^h \in \mathscr{H}^1(\Omega) ~\vert  ~u^h = ~\bar{u}\; \mathrm{on}\;  \Gamma^u\}$,  such that $\forall ~w^h \in \mathscr{V}^h \subset \mathscr{V}$, where $\mathscr{V}^h= \{ w^h \in\mathscr{H}^1(\Omega)~\vert  ~w^h = ~0 \;\mathrm{on}\;  \Gamma^u\}$, the finite-dimensional (Galerkin) weak form of the problem is satisfied. The variations $w^h$ and trial solutions $u^h$ are defined component-wise using a finite number of basis functions,
\begin{equation}
w^h = \sum_{a=1}^{n_\mathrm{b}} c^a N^a, \quad \qquad u^h = \sum_{a=1}^{n_\mathrm{b}} d^a N^a,
\label{eq:basisdef}
\end{equation}
\noindent where $n_\mathrm{b}$ is the dimensionality of the function spaces $\mathscr{S}^h$ and $\mathscr{V}^h$, and $N^a$ represents the basis functions. To obtain the Galerkin weak forms, we multiply each equation in strong form in (\ref{eq:S_2D}-\ref{eq:D_2D}) by a weighting function $w_\text{S}^h,w_\text{I}^h,w_\text{R}^h,w_\text{D}^h$, respectively, integrate by parts, apply boundary conditions appropriately, and use the Backward Euler method for time-discretization with $(\bullet)_n$ denoting a discretized quantity at time $t_n$ and $\Delta t$ being the time step. See Ref.~\cite{Wangetal-COVID2020} for details. This leads to:
\begin{align}
    \int_{\Omega}w^h_\text{S} \frac{S^h_{n} - S^h_{n-1}}{\Delta t}  \text{d}s &=-\int_{\Omega} \mathcal{D}_\text{S}\nabla w^h_\text{S}\cdot\nabla S^h_n\text{d}s -\int_{\Omega}w^h_\text{S}\left(\frac{\beta}{N} S^h_nI^h_n+\gamma R^h_n \right)\text{d}s\label{eq:S_2D_weak}\\
    \int_{\Omega}w^h_\text{I}\frac{I^h_n - I^h_{n-1}}{\Delta t}\text{d}s &=-\int_{\Omega}\mathcal{D}_\text{I}\nabla w^h_\text{I}\cdot\nabla I^h_n\text{d}s+\int_{\Omega}w^h_\text{I}\left(\frac{\beta}{N}S^h_nI^h_n-\mu I^h_n-\alpha I^h_n \right)\text{d}s\label{eq:I_2D_weak}\\
    \int_{\Omega}w^h_\text{R}\frac{R^h_n - R^h_{n-1}}{\Delta t}\text{d}s &=-\int_{\Omega}\mathcal{D}_\text{R}\nabla w^h_\text{R}\cdot\nabla R^h_n\text{d}s +\int_{\Omega}w^h_\text{R}\left(\mu I^h_n-\gamma R^h_n\right)\text{d}s\label{eq:R_2D_weak}\\
   \int_{\Omega}w^h_\text{D} \frac{D_n - D_{n-1}}{\Delta t}\text{d}s &=\int_{\Omega}w^h_\text{D}\alpha I^h_n\text{d}s\label{eq:D_2D_weak}
\end{align}
where the boundary terms vanish because we assume that the sub-populations do not leave the region--zero flux boundary conditions.

In our previous work~\cite{Wangetal-COVID2020}, we have characterized the coefficients to vary via a global-in-time polynomial basis. While the inferred model reproduced the trends, there was a notable error over time of the statewide sub-populations estimates $S(t,\bx), I(t,\bx), R(t,\bx), D(t,\bx)$ obtained by forward simulation with inferred quantities (See Figure 14 and 15 in \cite{Wangetal-COVID2020}). 

\begin{figure}
    \centering
    \subfigure[Michigan State]{\includegraphics[width=0.4\textwidth]{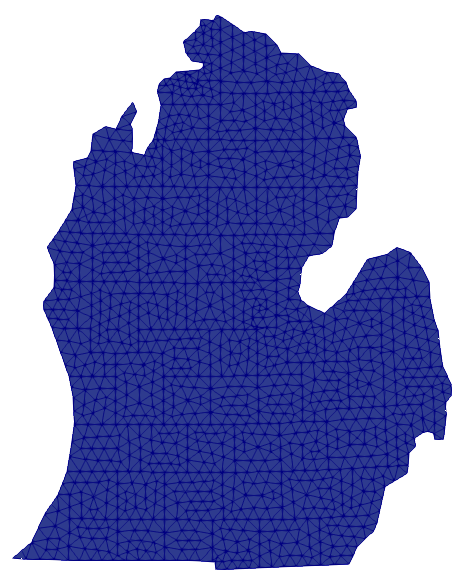}}
    \subfigure[Mexico]{\includegraphics[width=0.55\textwidth]{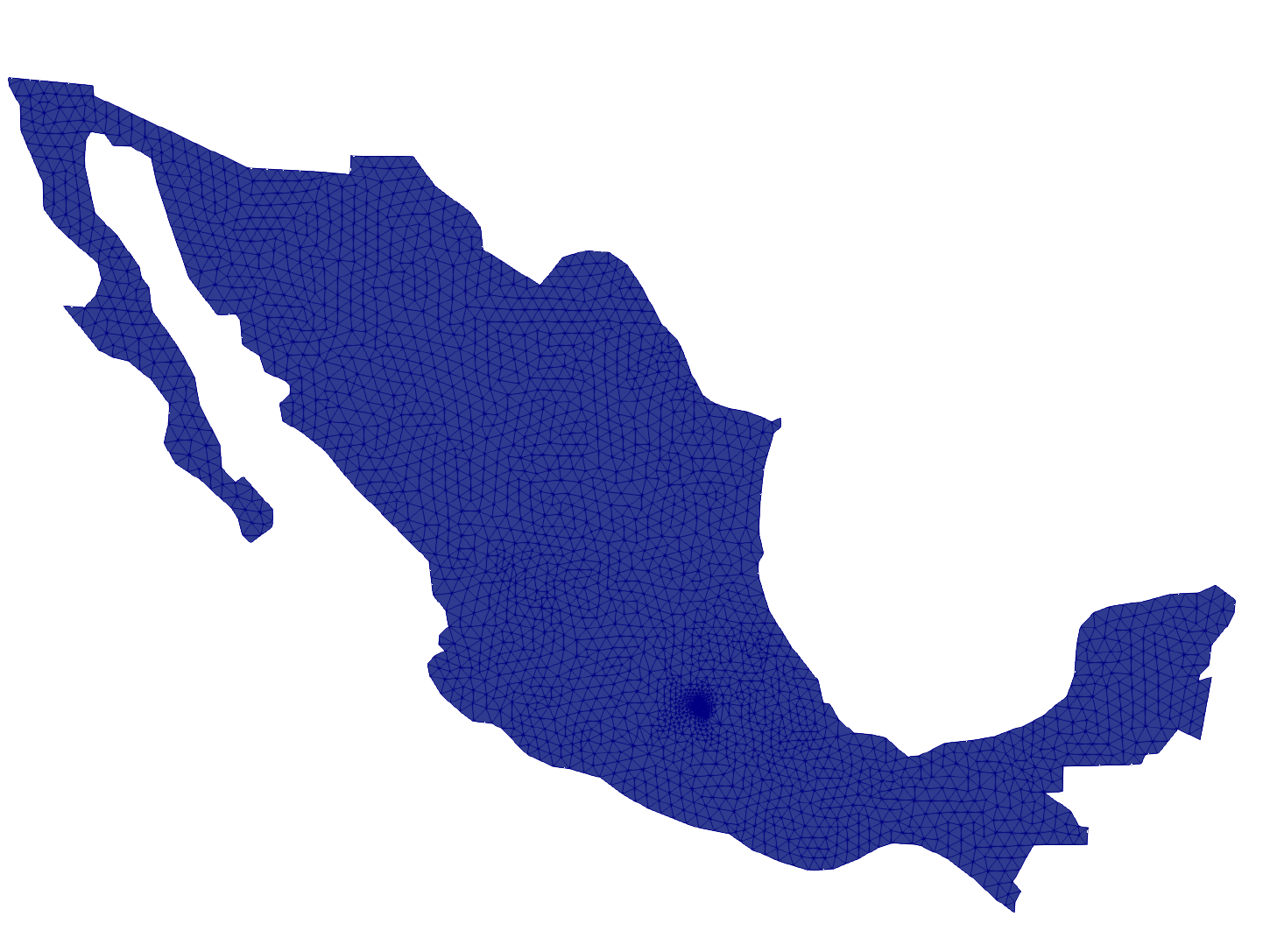}}
    \caption{Reconstructed maps of Michigan State and Mexico with finite element mesh. }
    \label{fig:mesh}
\end{figure}
Additionally, the highly complex geometry of the State of Michigan, and of Mexico (See maps in Figure \ref{fig:mesh}), and potentially highly nonuniform distributions of the coefficients in space makes it challenging to characterize the coefficients with simple basis functions. Global polynomials in space could not sufficiently resolve the emergence and disappearance of ``hot spots" and ``cold spots"~\cite{Wangetal-COVID2020}. In this communication, we allow the coefficients $\beta,\gamma,\mu,\alpha, \mathcal{D}_\text{S},\dots,\mathcal{D}_\text{R}$ of the PDE SIRD model to vary over space via finite-dimensional, locally supported representations as we do for the primary variables $S(t,\bx), I(t,\bx), R(t,\bx), D(t,\bx)$. Further more, we allow the coefficients to vary daily, leading to:
\begin{align}
\beta^h_n = \sum_{a=1}^{n_p} \beta^a_n N^a, \qquad \gamma^h_n = \sum_{a=1}^{n_p} \gamma^a_n N^a, \qquad \mu^h_n = \sum_{a=1}^{n_p} \mu^a_n N^a \qquad \alpha^h_n = \sum_{a=1}^{n_p} \alpha^a_n N^a
\label{eq:SIRDparamsinterp}
\end{align}
\begin{align}
\mathcal{D}_{\text{S}_n}^h = \sum_{a=1}^{n_p} \mathcal{D}_{\text{S}_n}^a N^a,\qquad \mathcal{D}_{\text{I}_n}^h = \sum_{a=1}^{n_p} \mathcal{D}_{\text{I}_n}^a N^a,\qquad \mathcal{D}_{\text{R}_n}^h = \sum_{a=1}^{n_p} \mathcal{D}_{\text{R}_n}^a N^a
\label{eq:SIRDdiffinterp}
\end{align}
where, as for the primary variables, the subscripts $(\bullet)n$ denote the coefficients on day $n$. With this, the PDE SIRD equations become:

\begin{align}
    \int_{\Omega}w^h_\text{S} \frac{S^h_{n} - S^h_{n-1}}{\Delta t}  \text{d}s &=-\int_{\Omega} \mathcal{D}^h_{\text{S}_n}\nabla w^h_\text{S}\cdot\nabla S^h_n\text{d}s -\int_{\Omega}w^h_\text{S}\left(\frac{\beta^h_n}{N} S^h_nI^h_n+\gamma^h_n R^h_n \right)\text{d}s\label{eq:S_2D_weak_inv}\\
    \int_{\Omega}w^h_\text{I}\frac{I^h_n - I^h_{n-1}}{\Delta t}\text{d}s &=-\int_{\Omega}\mathcal{D}^h_{\text{I}_n}\nabla w^h_\text{I}\cdot\nabla I^h_n\text{d}s+\int_{\Omega}w^h_\text{I}\left(\frac{\beta^h_n}{N}S^h_nI^h_n-\mu^h_n I^h_n-\alpha^h_n I^h_n \right)\text{d}s\label{eq:I_2D_weak_inv}\\
    \int_{\Omega}w^h_\text{R}\frac{R^h_n - R^h_{n-1}}{\Delta t}\text{d}s &=-\int_{\Omega}\mathcal{D}^h_{\text{R}_n}\nabla w^h_\text{R}\cdot\nabla R^h_n\text{d}s +\int_{\Omega}w^h_\text{R}\left(\mu^h_n I^h_n-\gamma^h_n R^h_n\right)\text{d}s\label{eq:R_2D_weak_inv}\\
   \int_{\Omega}w^h_\text{D} \frac{D_n - D_{n-1}}{\Delta t}\text{d}s &=\int_{\Omega}w^h_\text{D}\alpha^h_n I^h_n\text{d}s\label{eq:D_2D_weak_inv}
\end{align}
where the PDE SIRD parameters are interpolated from nodal variables as defined in Equations (\ref{eq:SIRDparamsinterp}) and (\ref{eq:SIRDdiffinterp}).
\section{System inference by field inversion using adjoint-based gradient optimization}
\label{sec:adjoint}
The system inference problem is to invert for the quantities $\beta^a_n,\gamma^a_n,\mu^a_n,\alpha^a_n, \mathcal{D}^a_{\text{S}_n},\dots,\mathcal{D}^a_{\text{R}_n}$. Since these quantities are interpolated via Equations (\ref{eq:SIRDparamsinterp}) and (\ref{eq:SIRDdiffinterp}) to be expressed as the corresponding fields $\beta^h_n,\gamma^h_n,\mu^h_n,\alpha^h_n, \mathcal{D}^h_{\text{S}_n},\dots,\mathcal{D}^h_{\text{R}_n}$ in (\ref{eq:S_2D_weak_inv}-\ref{eq:I_2D_weak_inv}), the system inference problems is one of field inversion. It is stated in Equations (\ref{eq:optim}-\ref{eq:loss}) as:

\begin{align}
\text{Given (\ref{eq:SIRDparamsinterp}-\ref{eq:SIRDdiffinterp}), at each}\; t_n:\quad\left(\beta^a_n,\dots,\mathcal{D}^a_{\text{R}_n}\right)_{a=1}^{n_p}&= \text{arg }\underset{(\beta^a_n,\dots,\mathcal{D}^a_{\text{R}_n}){a=1}^{n_p} }\min \text{ }  \ell_{i},\quad
    \text{such that (\ref{eq:S_2D_weak_inv}-\ref{eq:D_2D_weak_inv}) hold}
    \label{eq:optim}
\end{align}
and $\ell_{i}$ is the loss function defined:
\begin{align}
    \ell_{i}=&\int_\Omega W_\text{S}\left(S^h_n-S_n^\text{d}\right)^2+W_\text{I}\left(I^h_n-I_i^\text{d}\right)^2+W_\text{R}\left(R^h_n-R_n^\text{d}\right)^2+W_\text{D}\left(D^h_n-D_n^\text{d}\right)^2\text{d}v
    \label{eq:loss}
\end{align}
where $(\bullet)^\text{d}$ denotes data for the corresponding quantity. Due to the large differences in the magnitudes of different sub-populations, we choose the weights $W_\text{S},\cdots, W_\text{D}$ to be:
\begin{align}
    W_\text{S}=\frac{I_n^\text{d}} {\text{mean}\left(S_n^\text{d}\right) },\quad  W_\text{I}=\frac{I_n^\text{d}} {\text{mean}\left(I_n^\text{d}\right) },\quad  W_\text{R}=\frac{I_n^\text{d}} {\text{mean}\left(R_n^\text{d}\right) }, W_\text{D}=\frac{I_n^\text{d}} {\text{mean}\left(D_n^\text{d}\right) }.
\end{align}
The weights normalize the sub-populations and prioritize regions with higher infected populations. These regions are of greater interest for studying the progression of the disease as they tend to have a higher population density and, therefore, infected populations.

This PDE-constrained optimization problem is solved iteratively, and requires the gradient of the PDE constraint, Equations (\ref{eq:S_2D_weak_inv}-\ref{eq:D_2D_weak_inv}), with respect to parameters. We adopt classical adjoint-based gradient optimization. This approach involves a single linear solution of the adjoint equation of the original PDE constraint at each iteration, followed by solution of the fields to be inverted: $(\beta^h_n,\gamma^h_n,\mu^h_n,\alpha^h_n, \mathcal{D}^h_{\text{S}_n},\dots,\mathcal{D}^h_{\text{R}_n})$ and the updated forward solution $S^h_n, I^h_n, R^h_n, D^h_n$. In this work we use the L-BFGS-B optimization algorithm from \texttt{SciPy}\cite{2020SciPy} and the \texttt{dolfin-adjoint} software library \cite{dolfin-adjoint} to compute the gradient.


\section{Data preparation on maps of Michigan and Mexico}
\label{sec:data}
First, we constructed two-dimensional meshes for Michigan and Mexico that fully resolve the counties/states as shown in Figure \ref{fig:mesh}. The data are available as cumulative sub-population numbers $I^\text{d}_n, R^\text{d}_n, D^\text{d}_n$ at the county/state level. We used a uniform density of each sub-population to compute $I^\text{d}_n, R^\text{d}_n, D^\text{d}_n$ within the county/state, and applied Gaussian filtering to smooth the discontinuities at the county/state boundaries. Note that the discrete Gaussian filter can not be applied in a straightforward manner to unstructured meshes. Starting with a field $u$ that represents any of the four sub-population densities, and $G(\bx_0,\bx)=\frac{1}{2\pi\sigma^2}e^{-\frac{||\bx||^2}{2\sigma^2}}$ as the two dimensional Gaussian distribution function centered at $\boldsymbol{0}$ with standard deviation $\sigma$, which is related to the kernel size in the discrete Gaussian filter, we scale the filtered solution denoted by $u(\boldsymbol{x}_0)$ at each finite element node: 
\begin{align}
    u(\bx_0)=\frac{1}{{\int_\Omega G(\bx_0,\bx)\text{d}v}}{\int_\Omega G(\bx_0,\bx)u_\text{raw}(\bx)\text{d}v}
\end{align}
The spatio-temporal evolution of these fields was used in the system inference problem as described in Section \ref{sec:adjoint}.

\section{Results}
\label{sec:results}
\begin{figure}
    \centering
    \subfigure[Michigan]{\includegraphics[width=1\textwidth]{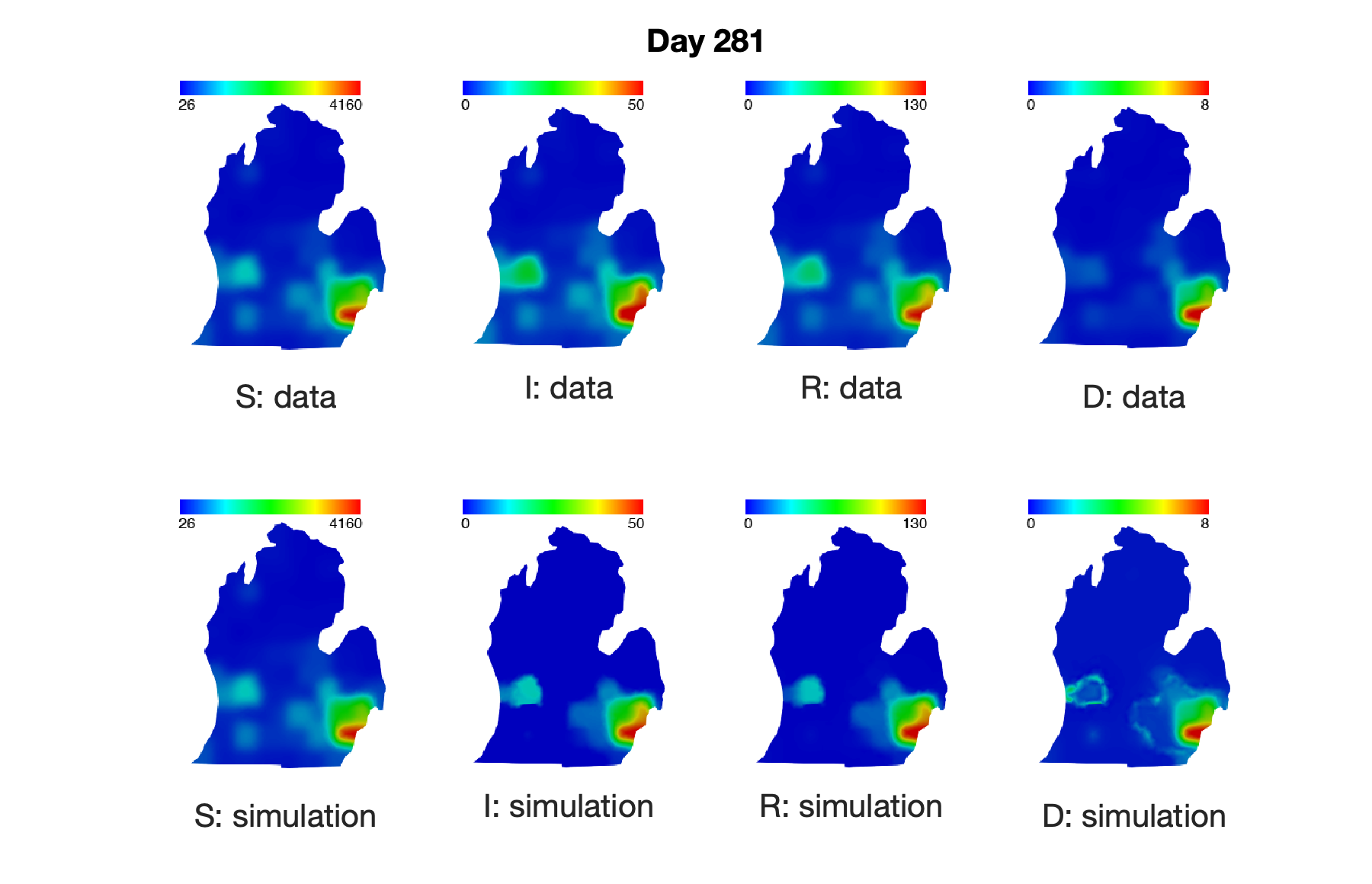}}
    \subfigure[Mexico]{\includegraphics[width=1\textwidth]{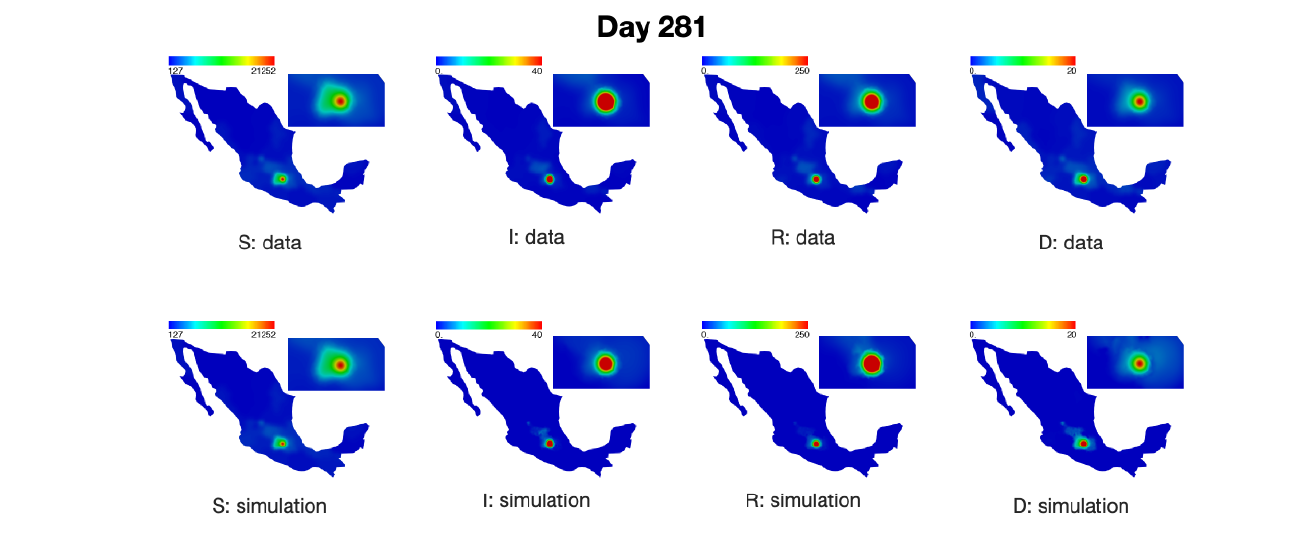}}
    \caption{Comparison of the data on distributions of the susceptible (S), infected (I), recovered (R) and deceased (D) sub-populations against forward PDE SIRD simulations with inferred quantities respectively. Data and simulation results are shown for Day 281 starting from March 23, 2020. The inset adjacent to the map of Mexico is a magnified view of the region surrounding Mexico City. The time-resolved dynamics may be seen as movies in Supplementary Information: michigan\_prediction.mp4 and mexico\_prediction.mp4.}
    \label{fig:simulation}
\end{figure}

Figure \ref{fig:simulation} shows the sub-populations $S(\bx,t), I(\bx,t), R(\bx,t), D(\bx,t)$ in both Michigan and Mexico obtained by forward simulation with inferred quantities compared with data on December 29, 2020 ($t=281$ days), where $t=0$ is March 23, 2020, the start of the lockdown in Michigan. The inferred model for Michigan accurately replicates the initial burst of disease and the following multiple waves around Detroit (please see the SI movie: michigan\_prediction.mp4). It also captures the second burst in the southwest of Michigan around the city of Grand Rapids. The high burden of the disease in these, the largest and second largest cities, respectively, in Michigan, reflects well-known socio-economic challenges related to Detroit in particular, and more generally reflected in other urban centers. Similarly, Mexico City, with highest population density in Mexico ($6,200/\text{km}^2$ \cite{Inegi-mex}), was the worst affected area in that country and dominated the evolution of the disease (See SI movie: mexico\_prediction.mp4).

\begin{figure}
    \centering
    \includegraphics[width=1\textwidth]{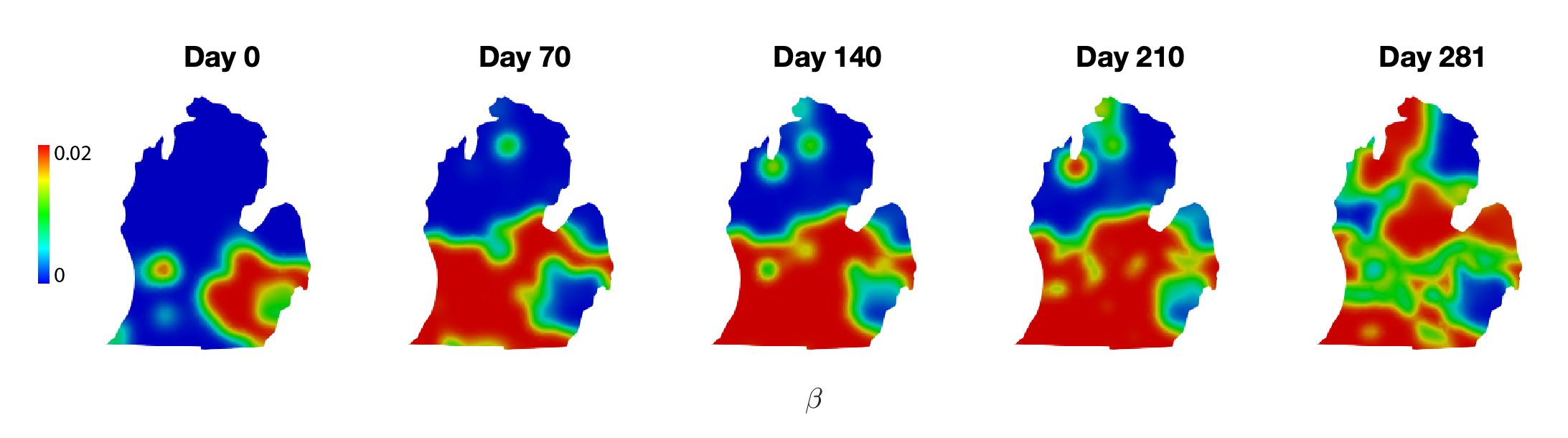}
    \includegraphics[width=1\textwidth]{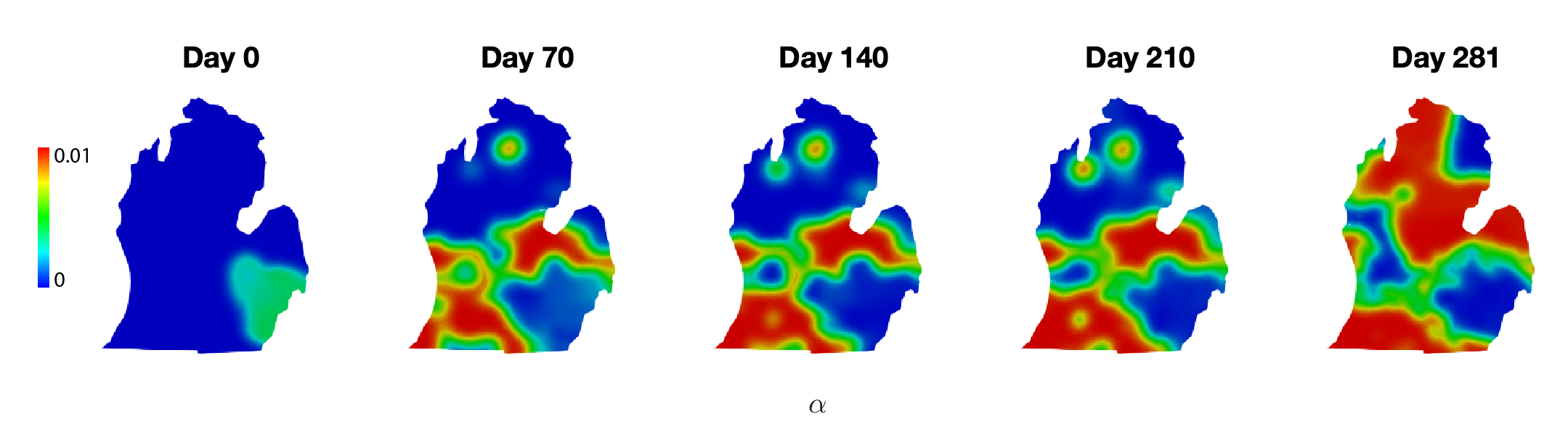}
    \includegraphics[width=1\textwidth]{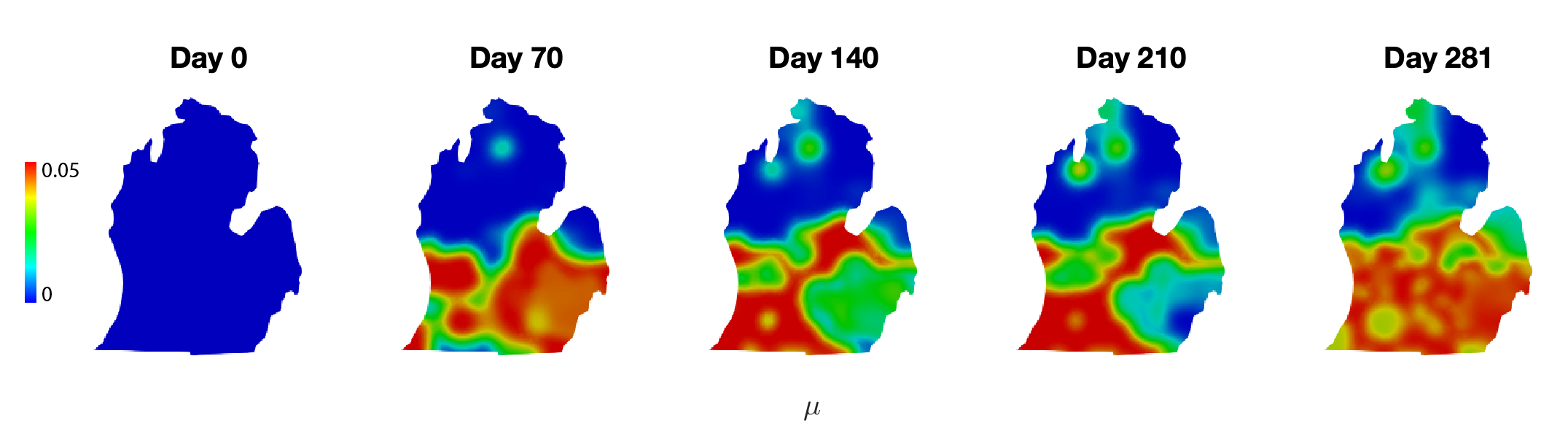}
    \caption{Inferred infection rate ($\beta$), death rate($\alpha$) and recovery rate ($\mu$) over the lower peninsula of Michigan starting from the lockdown on March 23, 2020. The time-resolved dynamics may be seen as movies  in Supplementary Information: michigan\_parameter.mp4. }
    \label{fig:michi_parameter}
\end{figure}

The low error between the simulation and data leads to greater confidence in the inferred parameters. Figure \ref{fig:michi_parameter} shows the inferred infection rate, death rate and the recovery rate in Michigan every 70 days starting from $t=0$ (the time-resolved dynamics are shown in SI movie: michigan\_parameter.mp4). The evolution of these inferred parameters reveals that the population's infection rate, $\beta(t)$, declined from the initially higher values in highly infected areas (such as Detroit), and spread to the western parts of Michigan. The death rate was mostly stable after May 2020 ($t>69$), and remained low in the more highly infected areas. This can be attributed to the ramp up of the public health campaign, hospitalizations and emergency response of the medical system, and prioritization to the more highly infected areas. The recovery rate around Detroit city evolved in multiple stages: increasing$\to$ decreasing $\to$ increasing, which was consistent with the multiple waves reflected in the data on the recovered population in this region (SI movie: michigan\_prediction.mp4).

\begin{figure}
    \centering
    \includegraphics[width=1\textwidth]{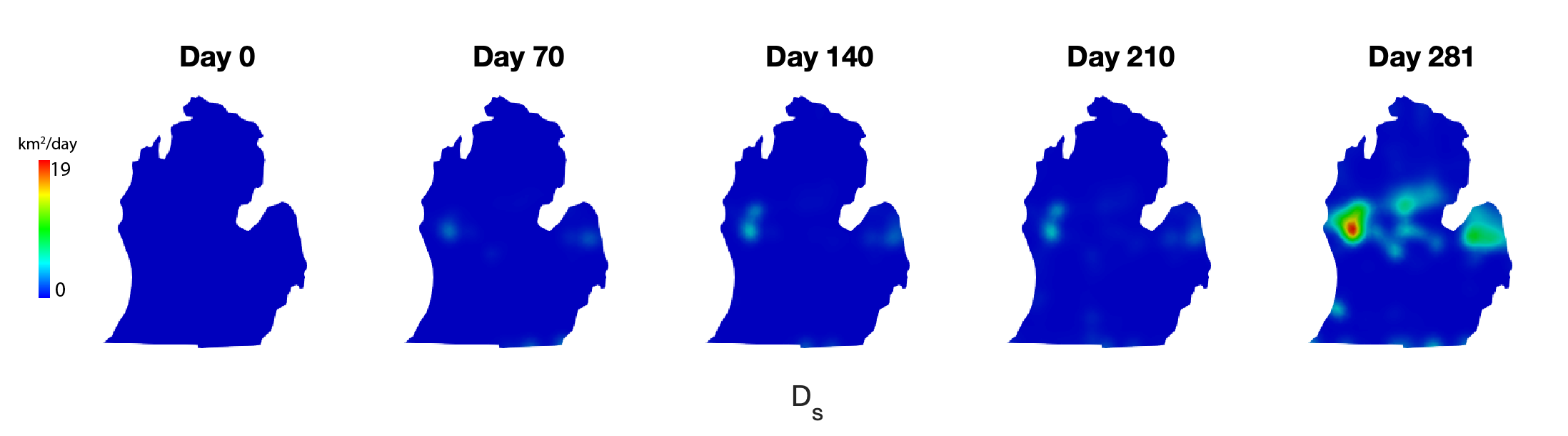}
    \includegraphics[width=1\textwidth]{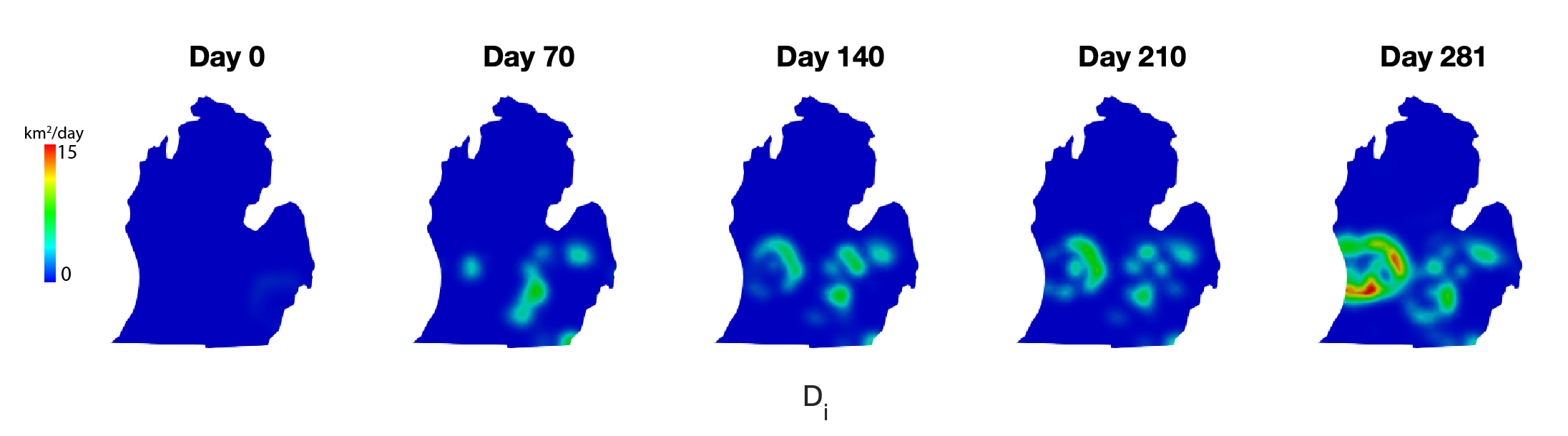}
   \includegraphics[width=1\textwidth]{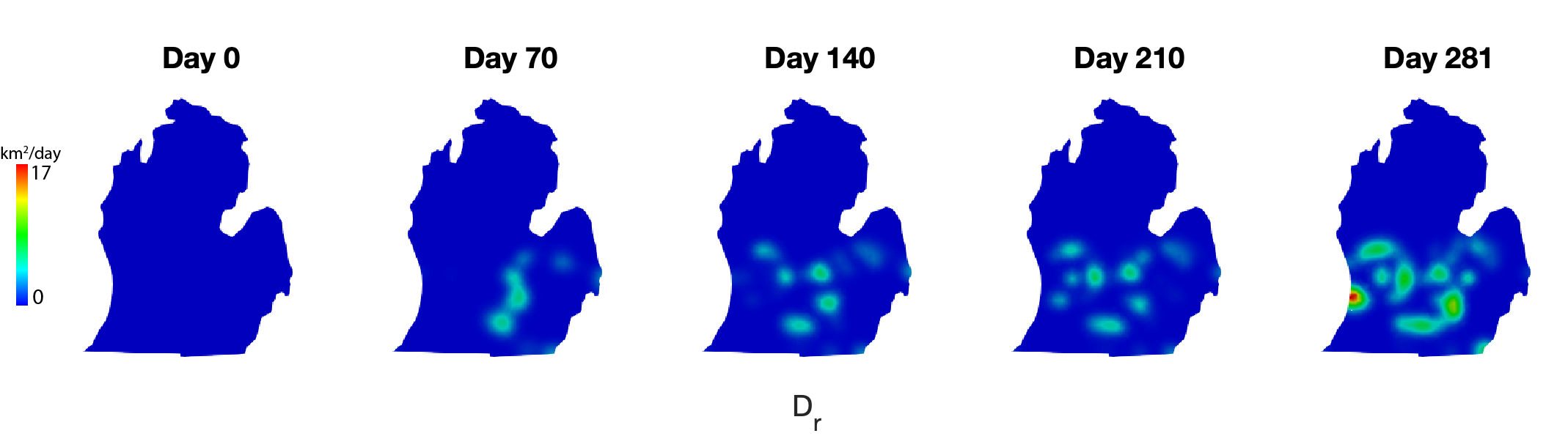}
    \caption{Inferred diffusivities of the susceptible, infected, and recovered sub-populations in Michigan starting from lockdown, March 23, 2020. The lower Peninsula of Michigan is about 446 km long from north to south and 314 km wide from east to west--scales that can help place the diffusivities in perspective. Time-resolved dynamics maybe seen as movies in Supplementary Information: mexico\_parameter.mp4.}
    \label{fig:michi_diffu}
\end{figure}

At the finest resolution, the mobility of the population during disease evolution may be approached via agent-based models refined to resolve individuals. However, given the difficulties encountered in effective contact tracing, and its acceptance by the population \cite{Zohdi2020,CHAN2021-contacttrac,KRETZSCHMAR2020contacttra}, an intriguing question to explore is whether simple reaction-diffusion models can detect the evidence of mobility in these data. Figure \ref{fig:michi_diffu} shows the inferred diffusivities of the susceptible, infected, and recovered sub-populations. Note that for field inversion, the population density data for each compartment was taken to be uniform within each county/state, since no finer grained information was available, and then subject to Gaussian smoothing before inference. Thus the density gradients, which drive the  inference of diffusivities,  arise at the counties/states scale more than they do at the intra-county/intra-state. Accordingly, the inferred diffusivities are meaningful on this scale. The lower Peninsula of Michigan is about 446 km long from north to south and 314 km wide from east to west--scales that can help place the diffusivities in Figure \ref{fig:michi_diffu} in perspective. The mobility of the infected population was always high around the highly infected areas. In Michigan, this infected population gradually shifted to the southwestern part of the state from the initial burst around Detroit. This finding is consistent with the second burst  around Grands during the evolution of the pandemic. The recovered population demonstrated a similar pattern of mobility, and was more active in the southern part of Michigan around the more highly infected regions. On the other hand, susceptible population closely tracks the total population. Since the  population at large has low mobility, the susceptible population's mobility is low in high population density areas. See SI movie michigan\_prediction.mp4 for these dynamics.

\begin{figure}
    \centering
    \includegraphics[width=1\textwidth]{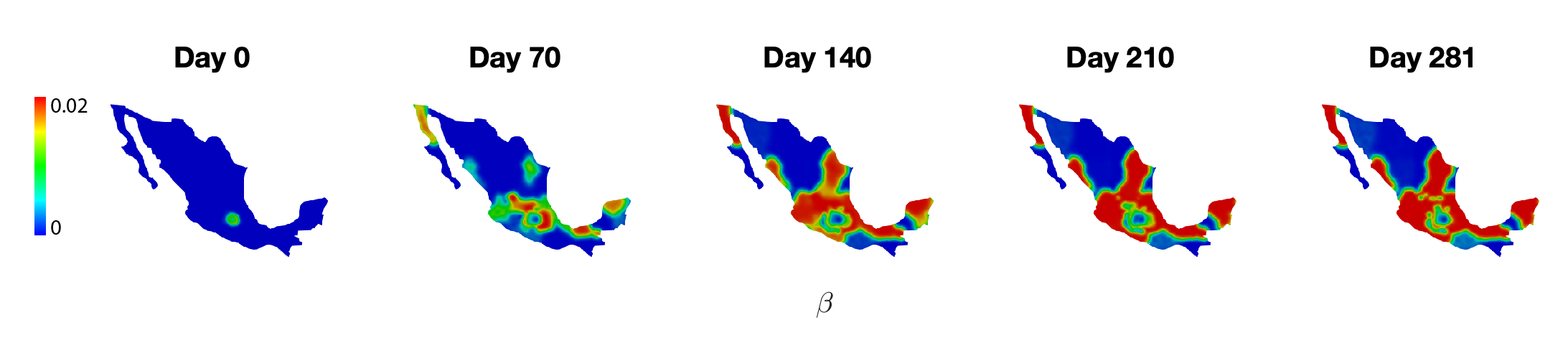}
    \includegraphics[width=1\textwidth]{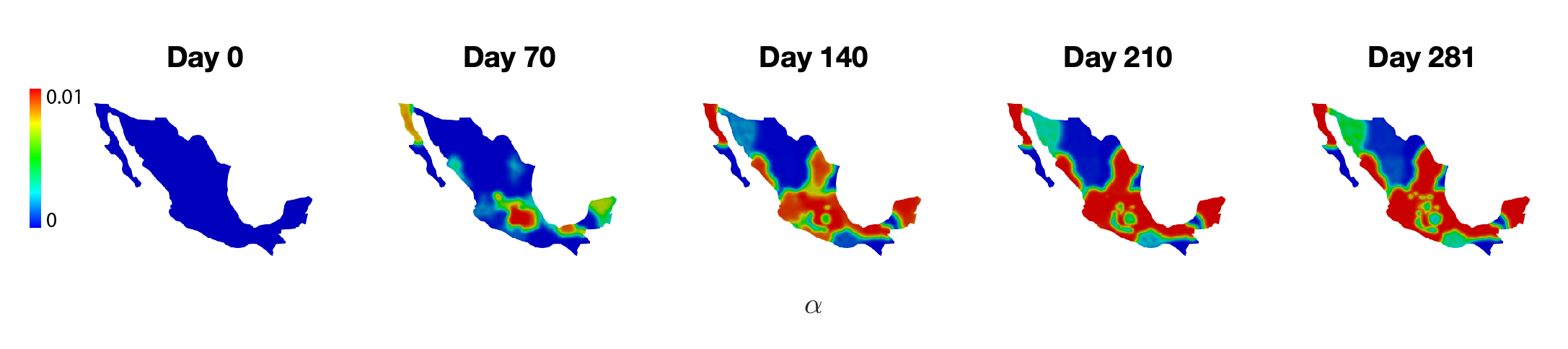}
    \includegraphics[width=1\textwidth]{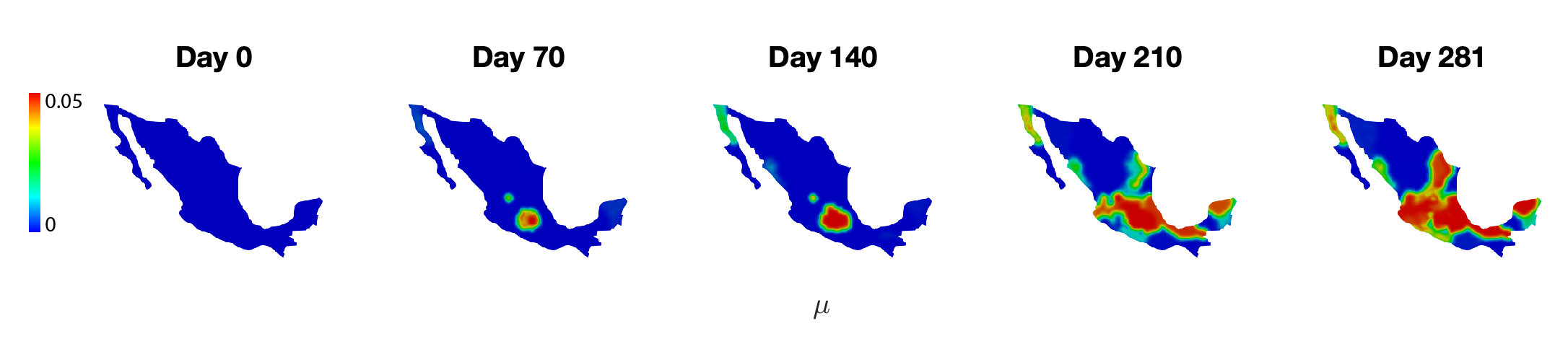}
    \caption{Inferred infection rate ($\beta$), death rate($\alpha$), recovery rate ($\mu$) in Mexico starting from lockdown, March 23, 2020. Time-resolved dynamics may be seen as movies in Supplementary Information: michigan\_parameter.mp4. }
    \label{fig:mexico_parameter}
\end{figure}

Figure \ref{fig:mexico_parameter} show the inferred infection rate, death rate and the recovery rate of the inferred model for Mexico. We can clearly see the spreading of the disease from Mexico City. Similar to the case of Michigan, infection rates, and to a lesser extent, death rates, were relatively lower in Mexico City, which is the most densely populated region of the country, than that in the surrounding cities. The recovery rate was high in Mexico city due to the relatively greater resources of the medical system there. The infection and death rates tended to be stable for five months following March 23, 2020, and the recovery rate gradually increased in more areas. Notably, far from the Mexico city, Baja California also displayed a high inferred rate of infection. We suspect this to be because it borders California, USA, and the international border restrictions did not contain the spread of the virus between the two regions. Unlike Mexico City, the death rate remained high, and the recovery rate did not increase to levels comparable to the capital, perhaps because of the looser restrictions in this popular tourist destination.

\begin{figure}
    \centering
    \includegraphics[width=1\textwidth]{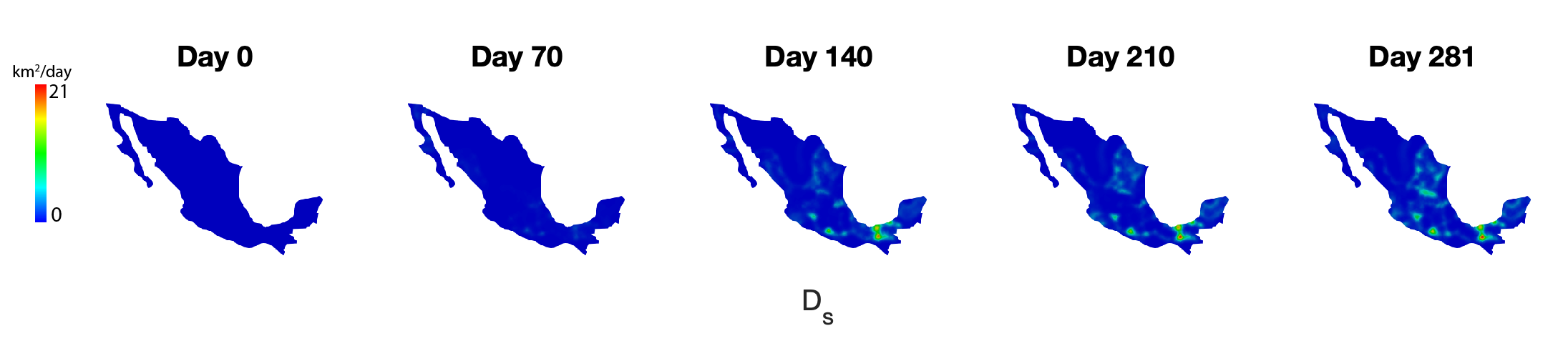}
    \includegraphics[width=1\textwidth]{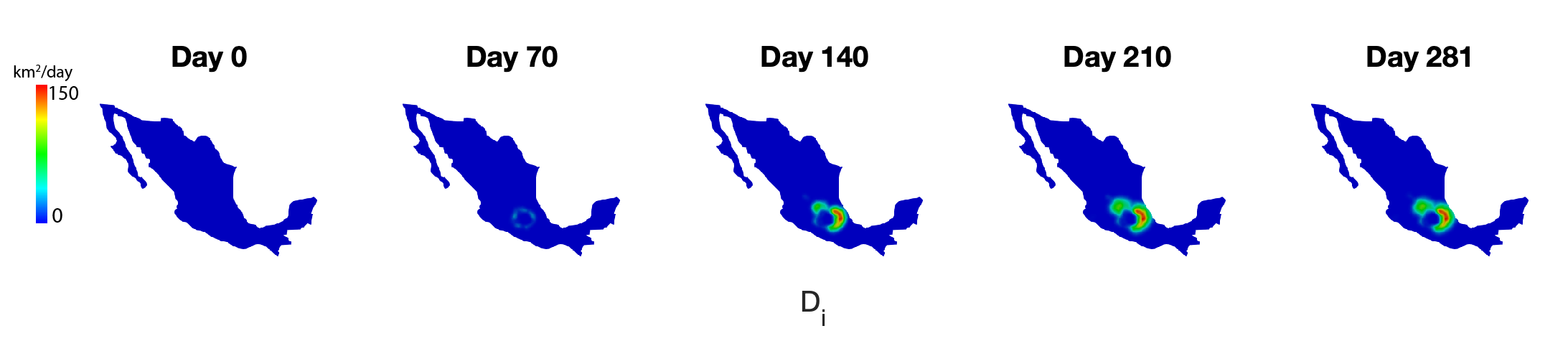}
   \includegraphics[width=1\textwidth]{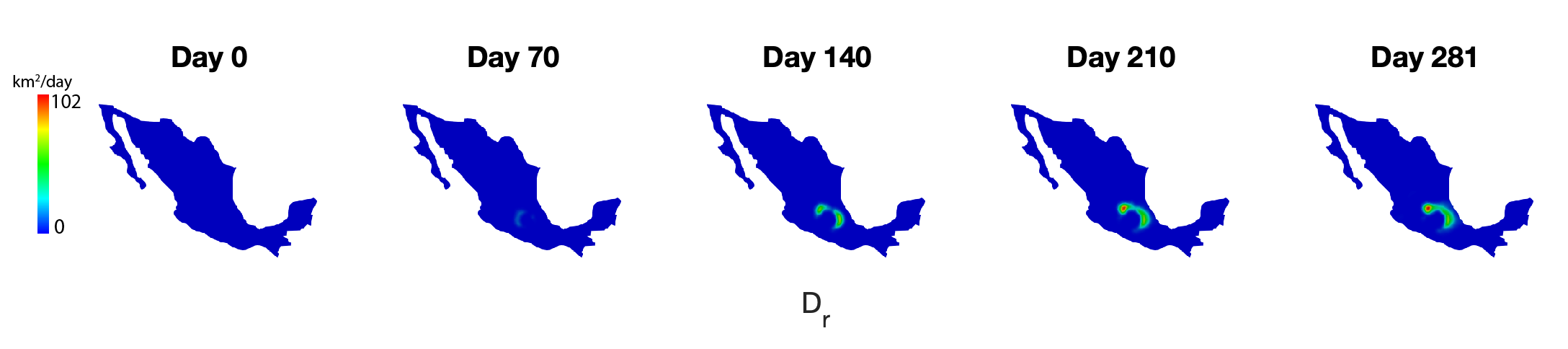}
    \caption{Inferred diffusivities of the susceptible, infected, and recovered sub-populations at day 281 at Mexico starting from lockdown, March 23, 2020. Mexico is about 3000 km long from north to south and 1900 km wide from east to west--scales that can help place the diffusivities in perspective. The time-resolved dynamics may be seen as movies in Supplementary Information: mexico\_parameter.mp4. }
    \label{fig:mexico_diffu}
\end{figure}

The diffusivities of the corresponding sub-populations of the inferred model for Mexico are shown in Figure \ref{fig:mexico_diffu}. Similar to Michigan, the mobility of infected and recovered sub-populations are higher around the highly infected Mexico City.  Mexico is about 3000 km long from north to south and 1900 km wide from east to west--scales that can help place the diffusivities in Figure \ref{fig:mexico_diffu} in perspective. Unlike the case in Michigan where there were multiple bursts in different cities, the mobilities of all sub-populations became stable after about 5 months from March 23, 2020. This may reflect differences in the proclivity toward domestic/local mobility of the populations of Michigan and Mexico--two regions with strongly contrasting social, economic and cultural characteristics.

\begin{figure}
    \centering
    \includegraphics[width=0.6\textwidth]{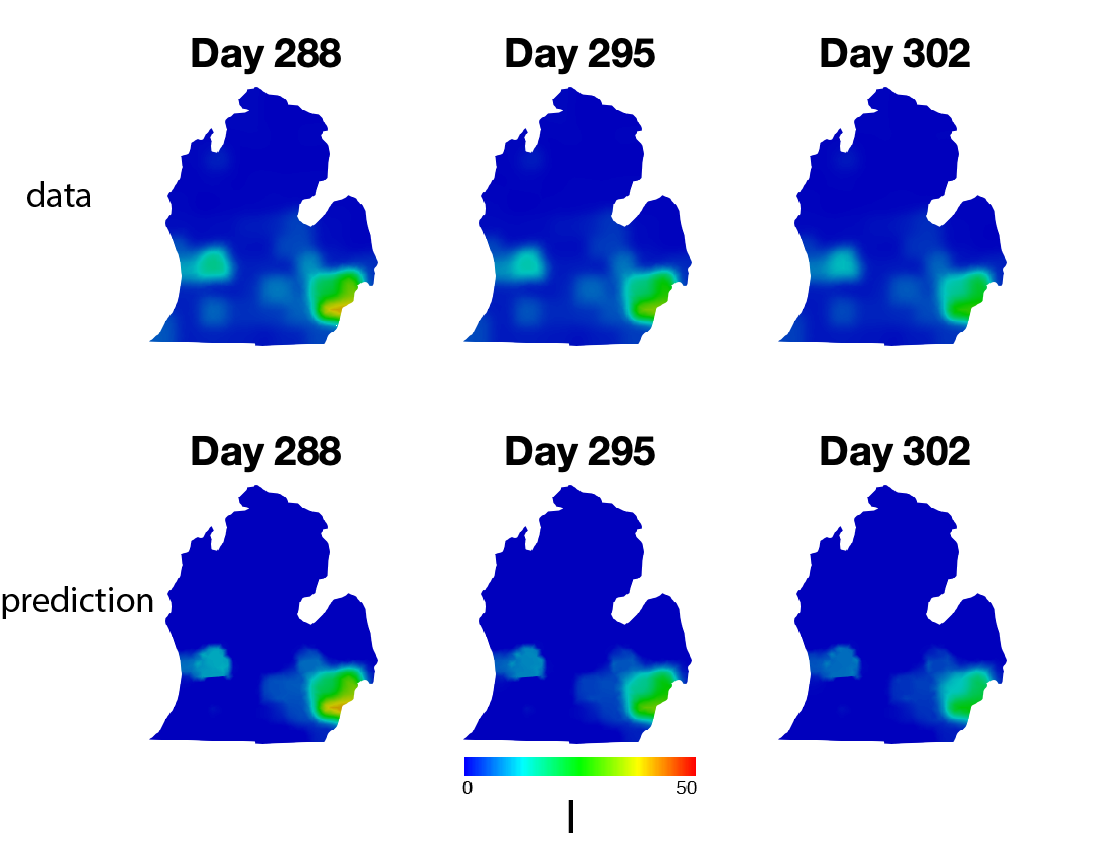}
    \caption{Prediction of infected population for 3 weeks against  the data for Michigan.}
    \label{fig:michi_future_pred}
\end{figure}

\begin{figure}
    \centering
    \includegraphics[width=1\textwidth]{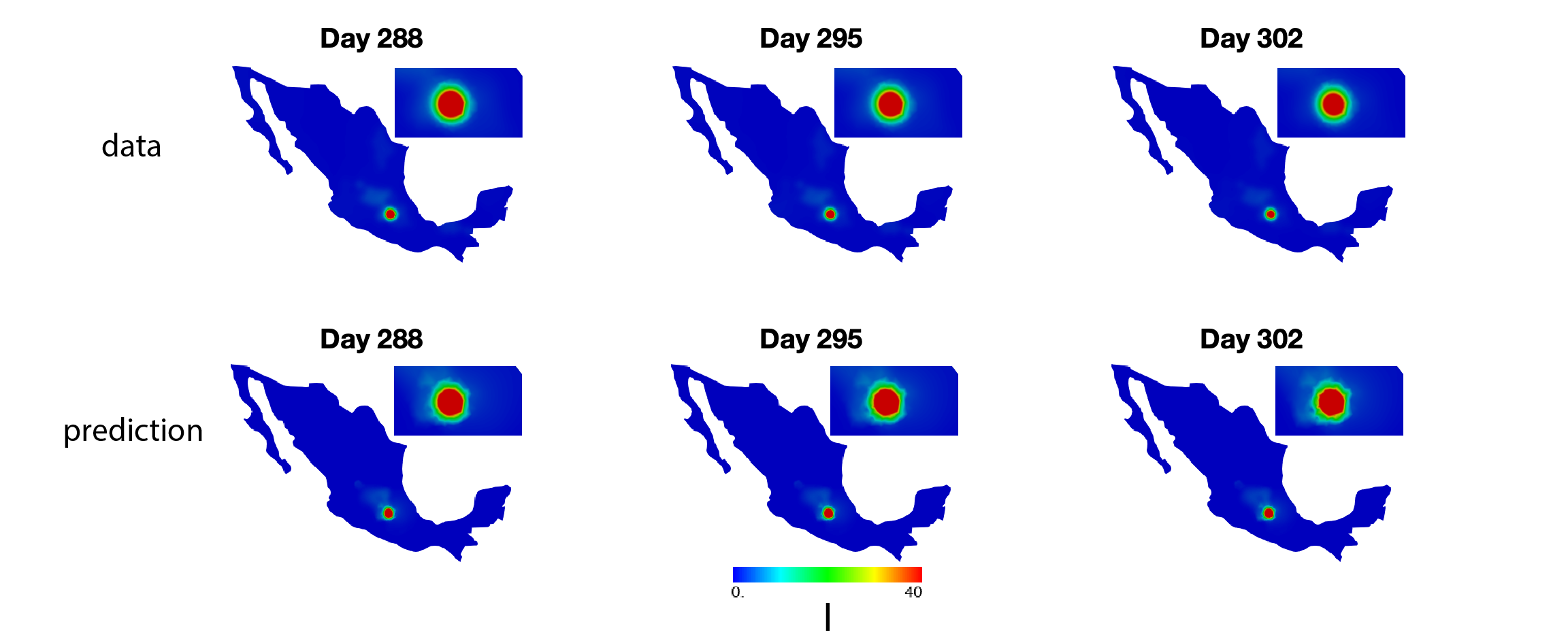}
    \caption{Prediction of infected population for 3 weeks against the data for Mexico. The inset adjacent to the map of Mexico is a magnified view of the region around Mexico City.}
    \label{fig:mexico_future_pred}
\end{figure}

Finally, taking the inferred parameters on the last day used for inference (Day 281), we predicted the evolution of sub-populations for three weeks (Days 282 to 303) using the inferred model. Figures \ref{fig:michi_future_pred} and \ref{fig:mexico_future_pred} show the predicted spatio-temporal evolution of the infected population against the raw data for both Michigan and Mexico. The inferred models captured closely the evolution of the infected-sub-populations, indicating that the dynamics of the disease tended to be steady in January 2021. The prediction of recovered and deceased sub-populations are shown in Figures \ref{fig:michi_future_pred_more} and \ref{fig:mexico_future_pred_more} under Supplementary Information. 

\section{Conclusion}
\label{sec:concl}

This communication builds upon our previous work~\cite{Wangetal-COVID2020} on system inference and machine learning from  data to study the progression of COVID-19 across the state of Michigan. We extended the PDE SIRD model by allowing the infection rate, death rate and the recovery rate, as well as the diffusivities of the susceptible, infected, and recovered sub-populations to vary over space and time. Using field inversion to infer the parameters as finite-dimensional fields on time scales of a single day, we obtained models to predict the evolution of disease with high accuracy. This provides us with the ability to analyze the dynamics of the disease through the inferred parameters, and make accurate predictions within a reasonable time frame. Particularly, we can detect the evidence of time and spatially varying mobility of the population through the simple diffusion-reaction models instead of the relying on the agent-based models which require individual's mobility data. The latter can prove challenging, technically as well as politically, to obtain.

As discussed in Section \ref{sec:results}, our inferred models capture the geographical spread of infection, the number of deaths and the size of the recovered population starting from one highly infected area to its surrounding cities and eventually spreading to further areas. Particularly, the higher infection and death rates in areas with low infection at later times suggests that more attention is needed in such locations. This may be due to a lack of medical services, or a lack of compliance with mitigation strategies. Our inferred models also reveals higher mobility surrounding the highly infected areas suggesting the importance of quarantine and social distancing. 

Finite-dimensional representation allows the parameters to accurately capture the spatial dependence, however the non-parametric representation makes the projection of these parameters beyond the data range extremely challenging. A prediction cannot be made with confidence if the dynamics of the disease reflected by these parameters are not stable. Of course, extrapolation is challenging in almost all data-driven methods. One possible alternate is to develope surrogate models of these parameters via time dependent neural networks under the constraints of the SIRD model to learn the spatial variation in time, and thus to make reasonable prediction of the dynamics in the evolution the disease, such as we have demonstrated previously~\cite{Wangetal-COVID2020}. Nevertheless, without including factors such as mobility restrictions or other mandates, only short time predictions may be accurate.

\section*{Acknowledgements}
We acknowledge the support of Defense Advanced Research Projects Agency (DARPA) under Agreement No. HR0011199002, ``Artificial Intelligence guided multi-scale multi-physics framework for discovering complex emergent materials phenomena''

\appendix
\section{Appendix: Additional results}
\begin{figure}[h]
    \centering
    \includegraphics[width=0.45\textwidth]{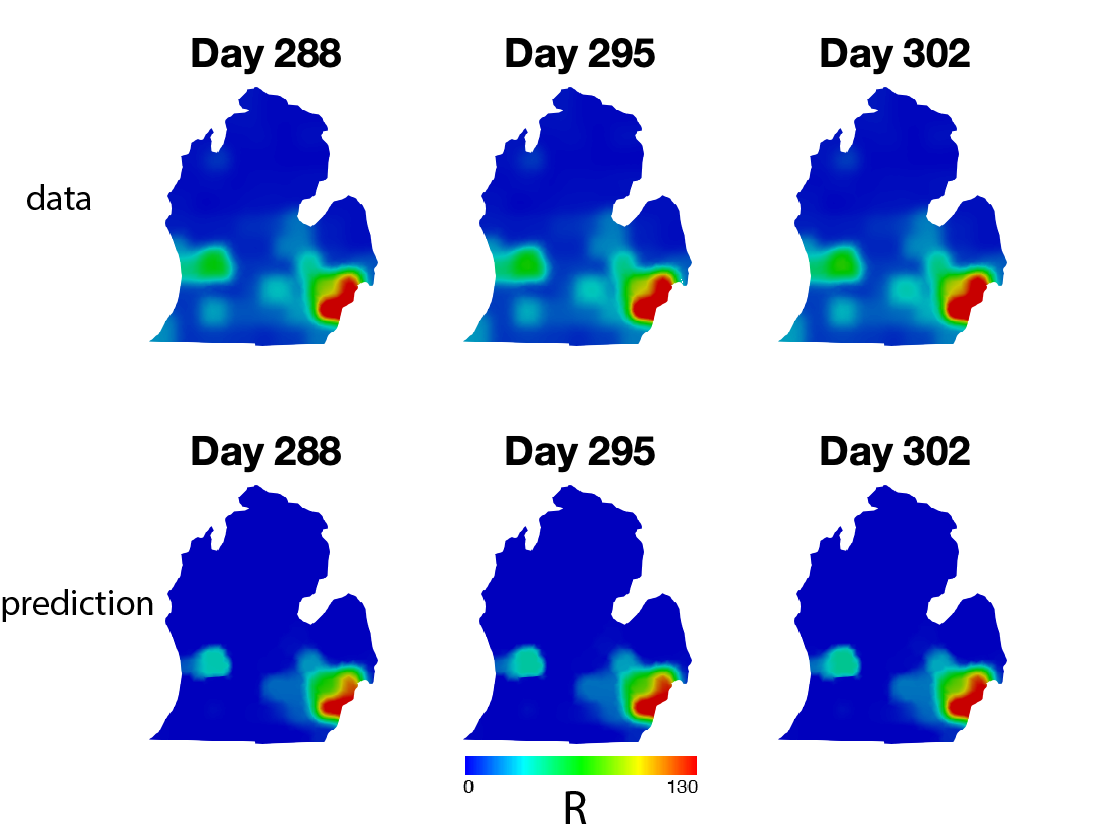}
    \includegraphics[width=0.45\textwidth]{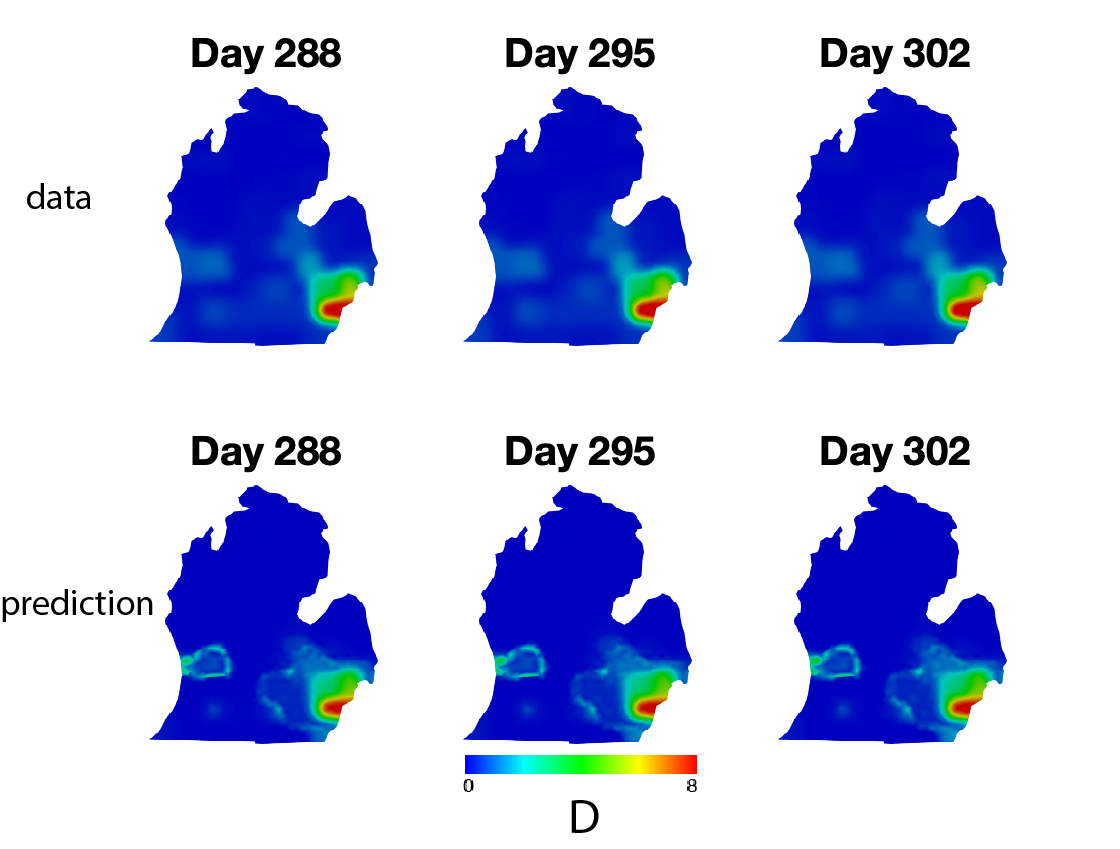}
    \caption{Prediction of the recovered and deceased sub-populations for 3 weeks against with the data at Michigan.}
    \label{fig:michi_future_pred_more}
\end{figure}

\begin{figure}[h]
    \centering
    \includegraphics[width=0.45\textwidth]{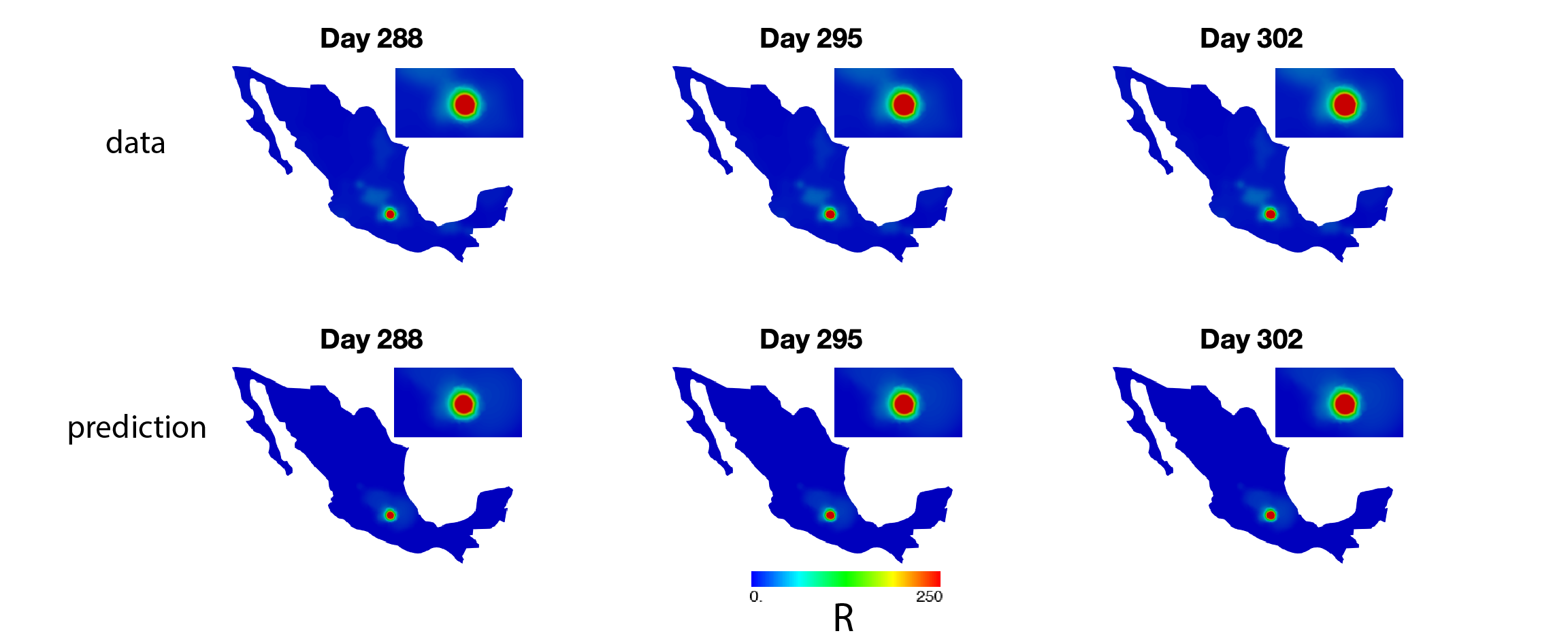}
    \includegraphics[width=0.45\textwidth]{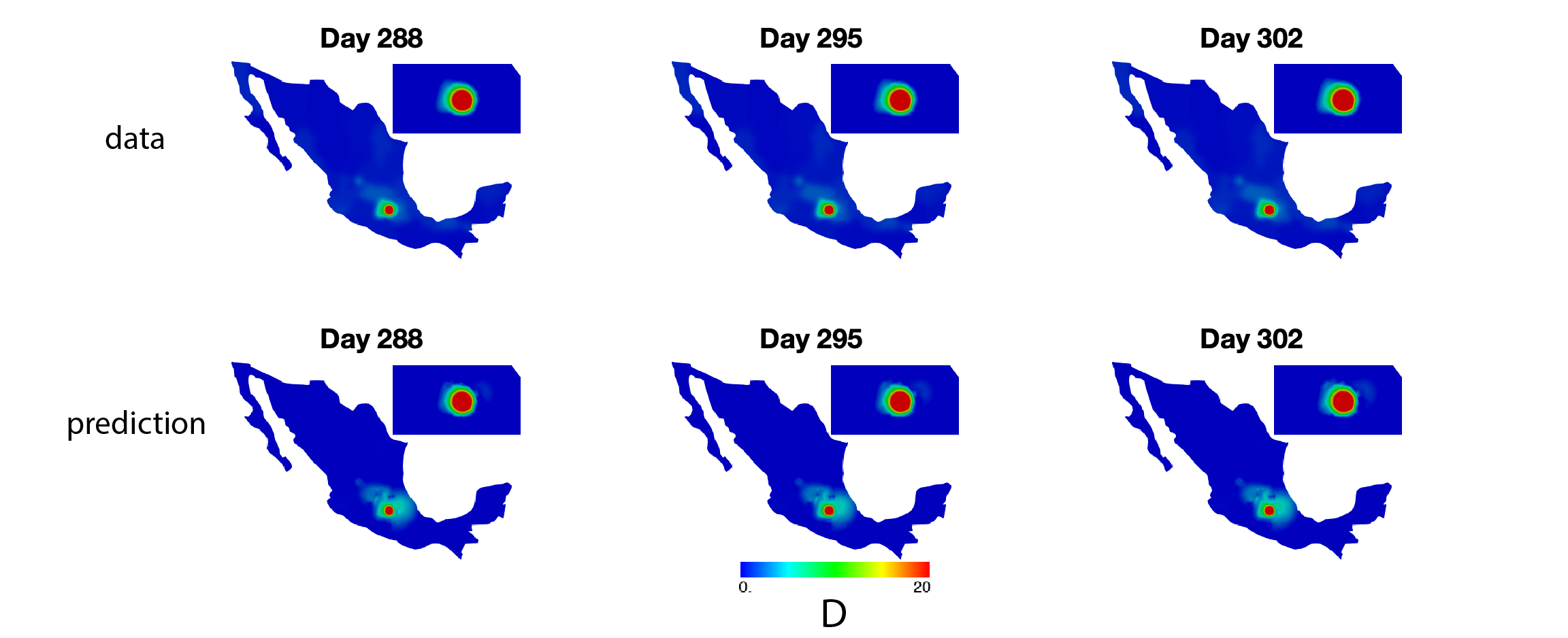}
    \caption{Prediction of the recovered and deceased sub-populations for 3 weeks against with the data at Mexico. The inset in the Mexico plot is a magnified view of Mexico City area.}
    \label{fig:mexico_future_pred_more}
\end{figure}

\bibliographystyle{unsrtnat}
\bibliography{ref.bib}
\appendix

\end{document}